\documentclass[oldversion]{aa}
\usepackage{graphicx}
\usepackage{txfonts}
\usepackage{natbib}
\usepackage{wrapfig}
\def\xmmp{2XMM1312\,}
\def\xmmplong{2XMMp J131223.4+173659}
\def\na1{Na{\sc I}}
\def\crat{s$^{-1}$}
\def\msun{M$_\odot$}

\begin{document}

   \title{The serendipituous discovery of a short-period eclipsing polar in
   2XMMp\thanks{Based on observations obtained with XMM-Newton, an ESA
   science mission with instruments and contributions directly funded by ESA
   Member States and NASA}}

   \author{J. Vogel\inst{1}
     \and K. Byckling\inst{2}
     \and  A. Schwope\inst{1}
     \and  J.P. Osborne\inst{2} 
     \and  R. Schwarz\inst{1}
     \and M.G. Watson\inst{2}}

   \institute{Astrophysikalisches Institut Potsdam,
                An der Sternwarte 16, 14482 Potsdam, Germany\\
              \email{jvogel@aip.de}
              \and
              Department of Physics and Astronomy, University of Leicester,
                Leicester LE1 7RH, UK
             }

\date{Received ; accepted }
\abstract{We report the serendipituous discovery of the new eclipsing
polar \xmmplong. Its striking X-ray light curve attracted immediate interest
when we were visually inspecting the source products of the 2XMMp catalogue. This
light curve revealed its likely nature as a magnetic cataclysmic
variable of AM Herculis (or polar) type with an orbital period of $\sim$
92\,min, which was confirmed by follow-up optical spectroscopy and
photometry. \xmmplong\, probably has a one-pole accretion geometry. It joins the
group of now nine objects that show no evidence of a soft component in
their X-ray spectra despite being in a high accretion state, thus escaping
ROSAT/EUVE detection. We discuss the likely accretion scenario, the 
system parameters, and the spectral energy distribution.}
\maketitle
\section{Introduction}
Magnetic cataclysmic variables (mCVs) of the AM Herculis type are
close binaries consisting of a magnetic white dwarf and a late-type
main-sequence star in synchronous rotation (see
\citealt{1995cvs..book.....W} for a comprehensive overview). The
late-type star loses matter via Roche-lobe overflow that initially
follows a ballistic trajectory and eventually is guided by strong
magnetic fields towards the polar regions of the white dwarf. The
accretion plasma cools via X-ray bremsstrahlung and optical cyclotron
radiation. A strong soft X-ray reprocessing component has been
observed in many systems. This key feature, together with strong
variability, gave the field a strong boost by the detection of about
45 new objects in EUVE/soft X-ray all-sky surveys
(e.g. \citealt{1995ASPC...85...99B}, \citealt{2002A&A...396..895S})
compared to about 20 known before ROSAT/EUVE. With no wide-angle, soft
X-ray survey telescope in space, new detections have become very
infrequent since the 90's.

The public release of the Sloan Digital Sky Survey opened a new
discovery channel. In a series of papers,
\citealt{2007AJ....134..185S} (and references therein) presented the
serendipituous CV content of the SDSS-sky. Among the several hundred
CVs, approximately a dozen magnetic objects were found. This small
fraction is probably more representative of the underlying
population. All the new systems are faint at optical and X-ray
wavelengths. A new class of low-accretion rate objects (formerly termed LARPs, \citealt{2002ASPC..261..102S}) 
in permanent low states was discovered by this route.
The implied mass accretion rates revealed wind accretion instead of
Roche-lobe overflow as the interaction channel, thus qualifying those
8 objects as magnetic pre-cataclysmic variables (\citealt{2005ApJ...630.1037S}, \citealt{2007A&A...464..647V}).

Since the launch of Chandra and XMM-Newton, several large-scale
optical identification programs have been started (e.g. Champ,
Champlane, AXIS, XBS). Not surprisingly, due to their small survey
area compared with ROSAT and EUVE, none of these led to the discovery
of any new magnetic CVs. All the public XMM-Newton observations have
been processed and surveyed by the XMM-Newton Survey Science Centre
(SSC).  The latest edition of the XMM catalogue, 2XMM, was made
available in August 2007 (Watson et al.~2007, in preparation), its
less elaborated predecessor, termed 2XMMp, was published in July
2006. The common feature of both catalogues is that the X-ray spectra
and light curves were prepared and published as standard source
products for objects with more than 500 source photons. Visual
inspection of all the light curves led to the serendipituous discovery
of \xmmplong~(for brevity \xmmp in the following).  The source stuck
out as a bright and prominently variable object showing periodic
modulations in its X-ray flux in the field of HD 114762, which is a
high-proper motion star. Optical follow-up observations secured the
tentative identification as an mCV.  We present the analysis of the
initial X-ray observations, and the spectroscopic and photometric
follow-up.
\section{Observations}
\begin{table}[t]
\caption{Log of X-ray and optical observations} 
\label{table:1}                        
\renewcommand{\footnoterule}{}
\begin{tabular}{lllll}       
\hline\hline
Date         &  Instrument  & Total Duration & Sampl.time\\
(Y/M/D)      &              & [ksec]         & [sec]\\
\hline
28/06/2004   &  XMM-PN      & 29.4           &\\
             &  XMM-MOS1    & 31.6           &\\
             &  XMM-MOS2    & 31.6           &\\
             &  XMM OM-UVW1 & 0.8            &\\
             &  XMM OM-UVM2 & 1.3            &\\
             &  XMM OM-UVW2 & 1.4            &\\
\hline
14/02/2007   &  CAHA 2.2m CAFOS   & 8.7      & 440\\
19/03/2007   &  CAHA 2.2m (R)     & 14       & 45\\

\hline
\end{tabular}\\
\end{table}
\subsection{XMM-Newton X-ray and UV observations}
XMM-Newton observed the field of HD 114762 (OBSID 0200000101) for
almost 32\,ks on June 28, 2004.  The observations were performed in
full window imaging mode with all three cameras through the medium
filter. All the X-ray and optical observations obtained are summarised
in Table \ref{table:1}. Routine processing for the production of 2XMMp
revealed the brightest X-ray source in the field at position RA, Dec
(2000) = 13h12m23.46s,$+17d36\arcmin59\fs5$, with an error of 0.36
arcsec. The source was detected at an off-axis angle of $\sim$ 6.3
arcmin. The three EPIC detectors detected more than 9500 photons from
the source corresponding to a mean count rate of $0.304\pm0.003$. The
standard products, i.e. the X-ray image, the DSS finding chart, the
X-ray spectra, and the X-ray light curves, were automatically
generated and visually inspected.

The PN X-ray light curve showed a striking behaviour with a periodic
on/off pattern on a time-scale of $\sim$ 90\,min. The peak count rate
reached 0.6\,\crat during the bright phase. Five bright phases were
covered by the observations. They were each disrupted by a short
($\sim$ 300 s) eclipse-like feature. The shape of the light curve in
general, its periodic behaviour, and the possible eclipse in the
bright phase, were all reminiscent of a synchronously rotating
magnetic CV (e.g. \citealt{2001A&A...375..419S},
\citealt{1988ApJ...328L..45O}), i.e., a so-called polar or \emph{AM
Herculis} star. In AM Her systems, there is an X-ray emitting
accretion region near one or both of the magnetic poles of the white
dwarf. The region may be eclipsed by the secondary and self-eclipsed
by the white dwarf, and thus the variability seen in the light curves
of AM Her systems can be due to the eclipsing of the emission
area. Prior to the eclipse the light curve shows a dip, a feature also
found in AM Her stars and caused by absorption of the softer X-rays
when the emitting accretion region becomes eclipsed by the accretion
stream. This striking light curve triggered a more detailed analysis
of the X-ray data and optical follow-up to secure the tentative
identification.

\begin{figure}
\centering
\includegraphics[width=6.7cm,bb=40 82 554 778,angle=-90,clip=]{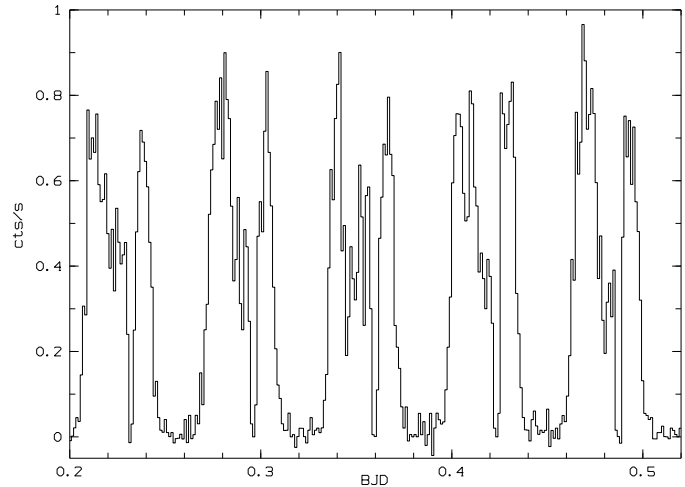}
\caption{Combined PN and MOS light curve with a binning of 110 s for the energy range 0.15-12.0 keV. Zeropoint is BJD = 2453185.0. For a phase folded light curve see Fig.\ref{caha_xmm_lc}}.
\label{f:lcori}
\end{figure}

We reanalysed the X-ray data obtained with XMM-Newton using SAS
version
7.0\footnote{http://xmm.vilspa.esa.es/external/xmm\_sw\_cal/sas.shtml}.
The usable energy range of the PN detector was extended down to 150 eV
using \emph{epreject}. Light curves and spectra were generated
according to the SAS Manual. \xmmp provided $\sim$ 6200 photons in the
EPIC PN detector, $\sim$ 2000 photons in the EPIC MOS1 detector and
$\sim$ 1700 photons in the EPIC MOS2 detector. The MOS count rate
reached $\sim$ 0.25\,\crat during the bright phase.  Nevertheless the
MOS data were checked for pile-up using \emph{epatplot} and were found
to be not affected. The background-corrected, binned X-ray light curve
of \xmmp, combining PN and MOS data, is reproduced in
Fig.~\ref{f:lcori}.  The same data, folded and averaged over the
orbital period of 91.85 min (see Sect. \ref{ephemeris}), are shown
together with the optical light curve in Fig.~\ref{caha_xmm_lc}.  The
mean X-ray spectrum is reproduced in Fig.~\ref{x_spec_fit} (we note
that \xmmp was not detected by ROSAT or by INTEGRAL).
\begin{table}
\caption
{\label{2xmmom}Count rates, fluxes and orbital phases of filter
observations with the Optical Monitor (OM) onboard XMM-Newton}
\renewcommand{\footnoterule}{}
\begin{tabular}{lcccc}
\hline\hline
Filter& $\lambda_{\rm eff}$    &  Countrate       & Flux ($10^{-16}$)          & Orbital phase\\
      & [\AA]   &                  & [erg cm$^{-2}$s$^{-1}$\AA$^{-1}$] &\\
\hline
UVW1  & 2910  & $0.264 \pm 0.045$   & $1.26 \pm 0.21$ & 0.25 - 0.40\\
UVM2  & 2310  & $0.099 \pm 0.022$   & $2.2 \pm 0.5$   & 0.97 - 0.21\\
UVW2  & 2120  & $0.039 \pm 0.019$   & $2.2 \pm 1.0$   & 0.52 - 0.77\\
\hline
\end{tabular}
\end{table}
The XMM Optical Monitor (OM) was operated in default imaging mode with
the U, UVW1, UVM2 and UVW2 filters. \xmmp was detected in the UVW1,
UVM2 and UVW2 filters, but not in U.  Since the OM imaging mode does
not allow for timing information, we get only an average flux for the
corresponding phase interval. This information is summarised in
Table~\ref{2xmmom}.
\subsection{Optical observations from Calar Alto}
\label{opticalobservations}
\begin{figure}
\centering
\resizebox{\hsize}{!}{
\includegraphics[clip=]{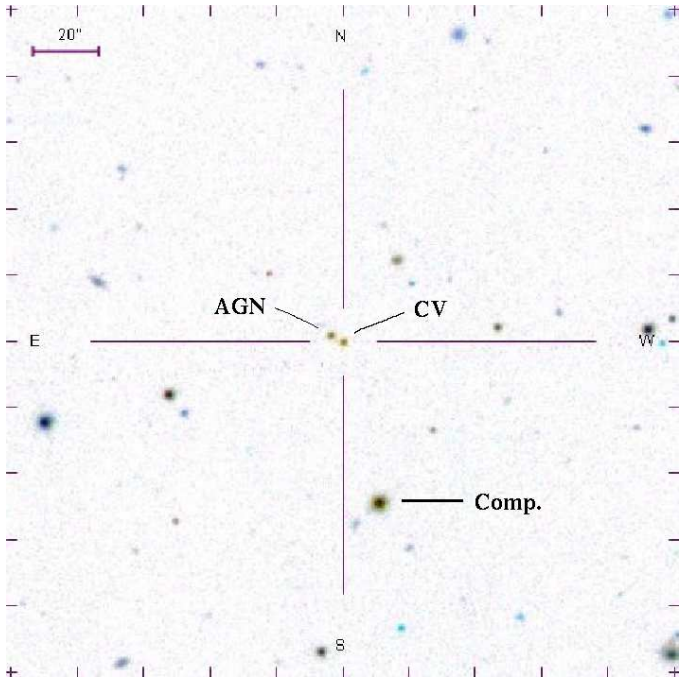}}
\caption{Finding chart of \xmmp\ created with the SDSS finding chart
tool. The new CV, the nearby AGN and the comparison star used for
differential photometry are marked}
\label{f:fc}
\end{figure}

The X-ray position of \xmmp\ was found to be coincident with the blue
object SDSS\,J131223.48+173659.1 having {\it ugriz} magnitudes of
19.66, 19.69, 19.76, 19.81, 19.83, respectively. A finding chart of
the field around the new CV is reproduced in Fig.~\ref{f:fc}.
\subsubsection{Time-resolved low-resolution spectroscopy}
Optical low-resolution spectroscopy was performed with the 2.2 m
telescope at Calar Alto on February~16, 2007 between UT
03:05--05:30. The telescope was equipped with the low-resolution
spectrograph and the CAFOS camera. The B-400 grism was used as
disperser resulting in spectra with full coverage of the optical range
from 3500--9500~\AA\ with a resolution of $\sim$ 28~\AA\ (FWHM). A
sequence of 20 spectra were taken with individual exposure times of 5
min each. Despite stable weather conditions, the object SDSS
J131223.75+173701.2 located 4.4 arcsec NE to the candidate ({\it
ugriz} = 20.72, 19.84, 19.67,19.59, 19.37) was put on the
spectrograph's slit to correct for slit losses and variable
seeing. The observations were accompanied by exposures of HgHeRb arc
lamp spectra for wavelength calibration and of the standard star BD
+75 325 for calibration of the instrumental response.

Data reduction was performed with
ESO-MIDAS\footnote{http://www.eso.org/sci/data-processing/software/esomidas/}.
A two-step procedure was applied to achieve spectrophotometric results
for the candidate CV. Firstly, a wavelength-independent correction
factor was derived from the slightly varying signal of the second
object on the slit and applied to the spectra of the target. Secondly,
the mean spectrum of the comparison object was folded through
SDSS-filter curves and differential magnitudes with respect to the
SDSS tabulated values computed. These were fitted with a second-order
polynomial as a function of wavelength and the correction function
applied to the spectra of the CV.

\begin{figure}
\centering
\resizebox{\hsize}{!}{
\includegraphics[clip=]{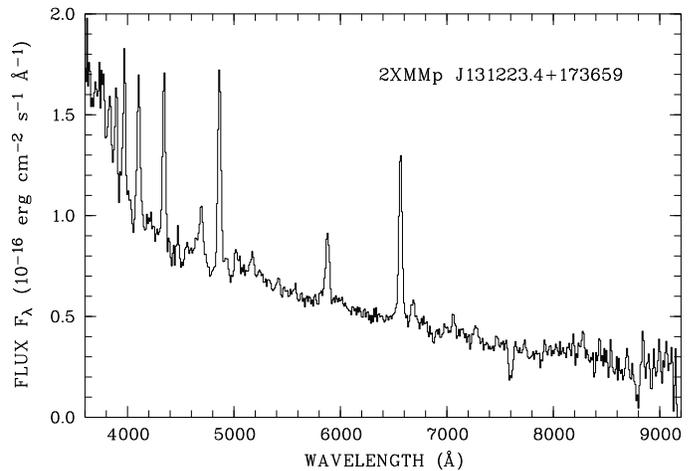}}
\caption{Average spectrum of \xmmp\ (= SDSS\,J131223.48+173659.1) obtained on
  February~16, 2007, with the CA2.2m telescope and CAFOS (total exposure 100 min).} 
\label{f:cvspec}
\end{figure}

\begin{figure}
\centering
\resizebox{\hsize}{!}{
\includegraphics[viewport=0 53 505 748,clip=]{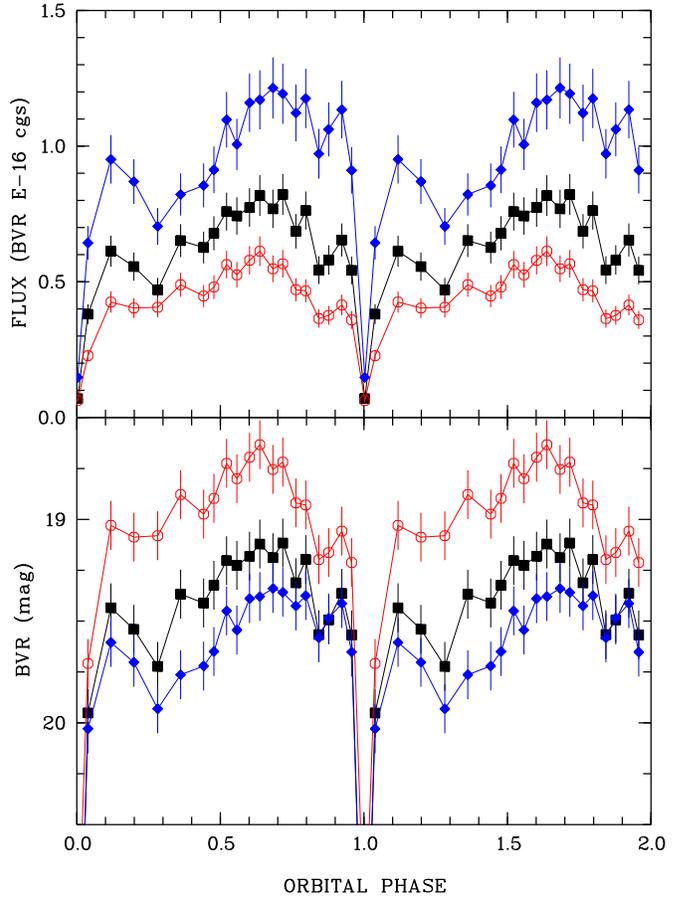}}
\caption{Optical light curves in BVR (in descending order for the upper plot and
ascending order for the lower plot) bandpasses determined from flux-calibrated
phase-resolved spectra of February 16, 2007. We estimated the errors from our
reduction procedure to $\sim$ 0.12\fm}
\label{f:lc_bvr}
\end{figure}

The mean spectrum of the CV candidate is reproduced in
Fig.~\ref{f:cvspec} and -- together with the X-ray light curve --
identifies \xmmp\ as a magnetic CV of AM Herculis type.  Approximate
$BVR$-band light curves of the CV were constructed by folding the
final spectra through the response curves of Bessel filters and are
reproduced in Fig.~\ref{f:lc_bvr}. Those light curves also show an
eclipse, there is one spectrum with a non-detection of the target
resulting in an upper limit to the brightness of the object of $V \sim
21\fm5$.

\begin{figure}
\centering
\resizebox{\hsize}{!}{
\includegraphics[clip=]{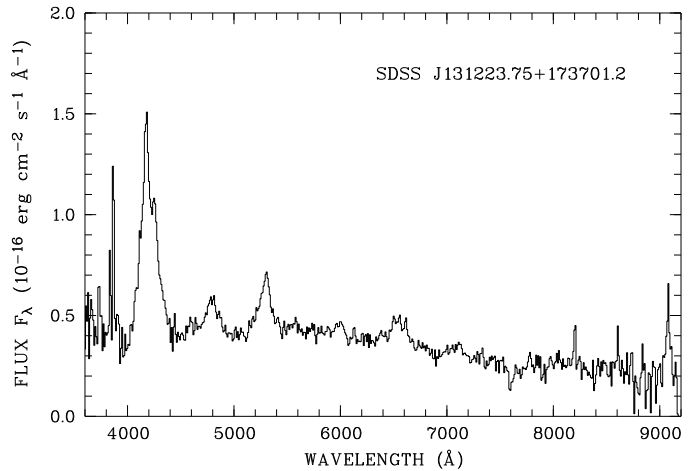}}
\caption{Average spectrum of the quasar SDSS J131223.75+173701.2 obtained on February~16, 2007.} 
\label{f:agnspec}
\end{figure}

The second object on the slit, which was used as a photometric
calibration source, turned out to be a quasar at redshift $z = 2.43$
with strong and broad emission lines of Ly$\alpha$, SI{\sc IV}, C{\sc
IV 1550}, and C{\sc III 1900}.  The mean spectrum of the QSO is
reproduced in Fig.\ref{f:agnspec}.

\subsubsection{Time-resolved differential R-band photometry}
\label{diffrband}
Optical R band photometry was performed one month after the
spectroscopy on March~19 between UT 00:32--04:25 with an exposure time
of 30 seconds at the Calar Alto 2.2m telescope. A total of 305
exposures were taken covering two and a half cycles of the binary and
three eclipses. The comparison used for photometric calibration is
marked in Fig.~\ref{f:fc}.

\begin{figure}
\includegraphics[width=8cm, bb = 50 50 525 738,clip=]{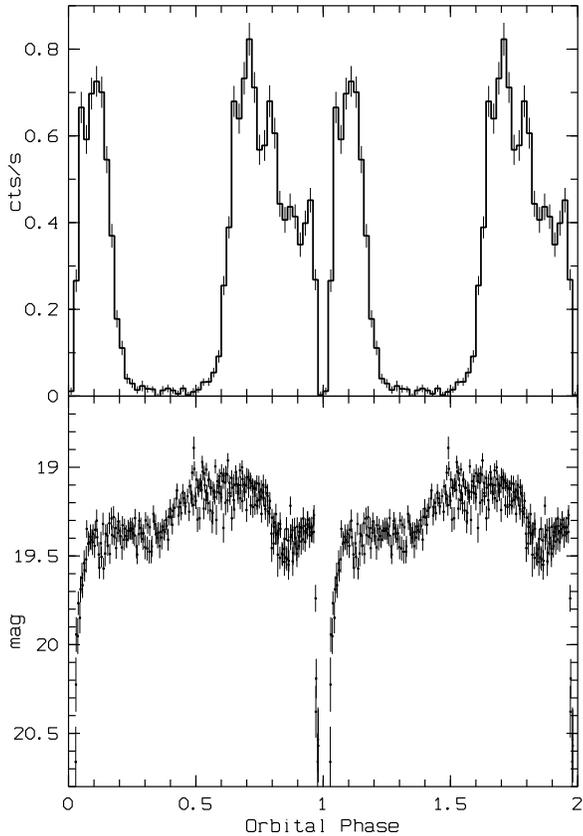}
\caption{The upper panel shows the phase folded X-ray light curve from 2004 composed 
        of EPIC PN and MOS counts with a binning of 0.02 phase units. The lower panel 
        shows the phase folded R band light curve from March 2007.}
\label{caha_xmm_lc}
\end{figure}

The phase-folded optical light curve of the CV is shown in
Fig.~\ref{caha_xmm_lc} together with the X-ray light curve (see
Sect. \ref{ephemeris} for the ephemeris). A zoomed-in version of the
same data centred on the optical eclipse is shown in
Fig.~\ref{caha_stream_ecl}. Similar to the X-ray light curve, the
overall light curve shows a bright hump followed by a dip prior to the
eclipse.

Out of the eclipse, the R-band brightness varied between $\sim 19\fm1$
and $\sim 19\fm4$. The CV was not detected in the individual exposures
at the bottom of the eclipse. A stacked image using those 11 images
was created. Running a source detection with
\emph{sextractor}\footnote{http://terapix.iap.fr} on this stacked
image revealed a source with an R band magnitude of $22\fm3 \pm 0\fm4$
exactly at the position of \xmmp. Since the source is just
1.2$\sigma$ above the background level and not free of problems due to
blending from the near AGN, we consider $22\fm3$ as the upper limit
for the brightness of the CV. Nevertheless, this might be the
detection of a polar secondary at the shortest orbital period so far
\citep{2006MNRAS.373..484K}. The orbital period of $\sim$ 92 min (see
next Section) implies a spectral type of M5.5$\pm$0.5
\citep{1998A&A...339..518B}. Using the absolute magnitudes from
\cite{1994AJ....107..333K} this results in a minimum distance of
$\sim$ 900 pc for a M5 and $\sim$ 350 pc for a M6 spectral type.
\section{Analysis and discussion}
\subsection{A combined X-ray/optical ephemeris}
\label{ephemeris}
We found an eclipse in all three datasets obtained so far. This
excludes anything else than an eclipse proper by the mass-donating
secondary star. Here we make an attempt to combine the eclipse epochs
from the three sets into one binary ephemeris. For the analysis of
the X-ray data a combined light curve of all the three EPIC cameras
with a binning of 5 seconds was used. The X-ray eclipse shows just the
ingress and egress of the accretion spot and does not resolve any
detail. The individual X-ray eclipse timings with an uncertainty equal
to the bin size were used to determine the times of the
mid-eclipses. The mean X-ray eclipse length resulting from the five
individual eclipses is 329$\pm$3 s.

The exposure time of the optical spectra was 300 seconds, but the last
spectrum in our sequence of 20 did not yield a significant
detection. Since the length of the X-ray eclipse is slightly longer
than this exposure time, we assume that this spectrum is
centred on the X-ray eclipse with the rather small uncertainty (compared
to the exposure time) of 60 seconds only.

The mean eclipse light curve from the R band photometry (see
Fig.~\ref{caha_stream_ecl}) clearly shows the ingress and the egress
of the white dwarf and/or the spot and the accretion stream. Since the
individual eclipses do not resolve these details, we used the mean
eclipse light curve -- phase folded with the period derived from the
X-ray data (see below) -- to determine the flux level for mid-ingress
and mid-egress of the white dwarf/spot. For the individual eclipses we
interpolated between single exposures to get the timings for the
chosen flux level, and thus the times of the ingress and egress of the
spot.

The five consecutive X-ray eclipses were used to derive a preliminary
period of 0.06378(3) d that could be used to connect the eclipse times
obtained from spectroscopy and photometry at Calar Alto without cycle
count alias. A linear regression to the four optical eclipse times
yielded a period of 0.063785(1) d.  This turned out to be sufficiently
accurate to connect the 2004 XMM-Newton X-ray data with the optical
data from 2007, again without cycle count alias. A linear regression
using all nine eclipse epochs revealed the finally accepted eclipse
ephemeris of:

\begin{equation}
\mbox{BJD(UT)} = 2453185.23204(4) + E \times 0.06378527(1)
\label{e:eph}
\end{equation}
which corresponds to an orbital period of P$_{orb}$ =  91.85079(1) minutes.
Numbers in parentheses indicate the uncertainties in the last digits. All
phases in this paper refer to the linear ephemeris of Eq.~\ref{e:eph}. The
mid-eclipse times for all the eclipses are given in Table \ref{2xmm_ecltimes}.

\begin{table}[t]
\caption{Mid-eclipse timings} 
\label{2xmm_ecltimes}                        
\renewcommand{\footnoterule}{}
\begin{tabular}{llrr}       
\hline\hline
Observation    & Time [BJD]        & Cycle & O-C [sec]\\ &&&\\
\hline
XMM EPIC 2004  & 2453185.23203(8) & 0     & -0.6\\
XMM EPIC 2004  & 2453185.29585(8) & 1     & 2.5\\
XMM EPIC 2004  & 2453185.35959(8) & 2     & -1.7\\
XMM EPIC 2004  & 2453185.42345(8) & 3     & 4.3\\
XMM EPIC 2004  & 2453185.48715(8) & 4     & -2.3\\                    
CAHA Spec 2007 & 2454146.7313(7)  & 15074 & 8.6\\
CAHA Phot 2007 & 2454178.5600(4)  & 15573 & -4.3\\
CAHA Phot 2007 & 2454178.6240(4)  & 15574 & 14.3\\
CAHA Phot 2007 & 2454178.6875(4)  & 15575 & -10.4\\
\hline
\end{tabular}\\
\end{table}
\subsection{The X-ray spectrum}
\label{xspectrum}
\begin{figure}
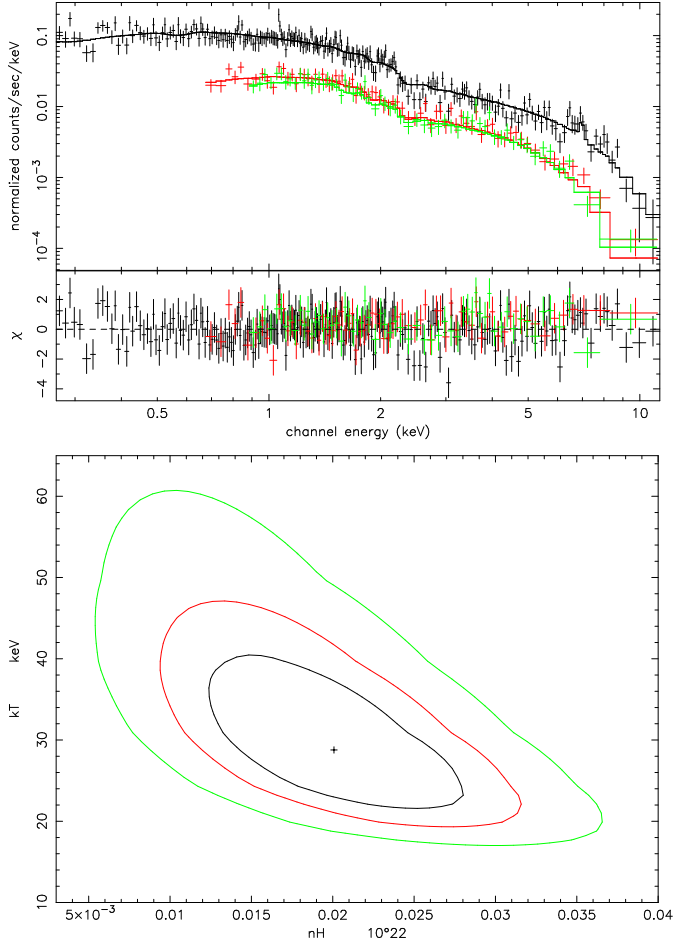

\includegraphics[width=6.0cm,bb=112 46 564 718,angle=-90,clip=]{9341fig6.ps}
\includegraphics[width=6.5cm,bb=79 46 570 720,angle=-90,clip=]{9341fig7.ps}
\caption{Spectral fit of the combined PN and MOS mean spectra with an absorbed MEKAL model, 
  together with the confidence range (0.68, 0.90 and 0.99) as a function of the MEKAL 
  temperature and the 
  column density of the interstellar absorption. }
\label{x_spec_fit}
\end{figure}
The PN and MOS X-ray spectra were extracted according to the SAS
manual. The data were binned to at least 20 cts/bin. The spectral fits
were generated using
Xspec\footnote{http://heasarc.gsfc.nasa.gov/docs/xanadu/xspec/}. A
good fit to the PN and MOS mean spectra (see Fig.~\ref{x_spec_fit})
was obtained with an absorbed MEKAL model with $\chi^2_{\nu}$=1.01,
and resulted as a plasma temperature of kT = 29(8) keV and a column
density of $N_{\rm{H}} = 2.0(6) \times 10^{20}$ cm$^{-2}$.  As a
comparison, a Galactic column density of $N_{\rm{H}} = 1.87 \times
10^{20}$ cm$^{-2}$ was obtained with the \emph{nh} program, which is
provided by the FTOOLS
package\footnote{http://heasarc.gsfc.nasa.gov/docs/software/ftools}.
The target coordinates correspond to a galactic latitude of
79$^{\circ}$, so this value seems still reasonable with respect to the
estimated distance.  The unabsorbed flux is $1.4 \times
10^{-12}$\,erg\,s$^{-1}$\,cm$^{-2}$ in the energy range
0.15--12.0\,keV. We also experimented with a more complex absorption
component using the \emph{pcfabs} model in Xspec, but in this model
the parameters of the absorber were not constrained by the data. The
above model parameters results in a predicted ROSAT count rate of 0.03
ct/s, slightly below the RASS detection limit of 0.05 ct/s, which
explains the non-detection during the RASS if we assume a comparable
accretion state for the XMM and the ROSAT observation and even if we
assume that the RASS scans covered the bright phase.
\begin{figure}
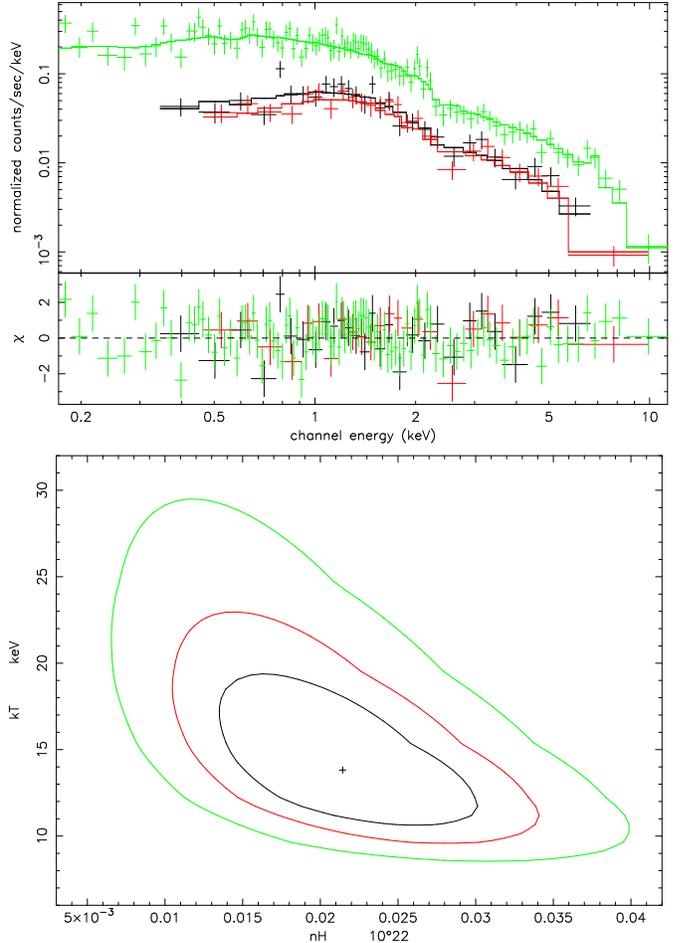

\includegraphics[width=6.0cm,bb=112 43 560 709,angle=-90,clip=]{9341fig8.ps}
\includegraphics[width=6.5cm,bb=79 45 567 718,angle=-90,clip=]{9341fig9.ps}
\caption{Spectral fit of the combined PN and MOS spectra of the X-ray bright phase before 
  the absorption dip with an absorbed MEKAL model, together with the confidence range 
  (0.68, 0.90 and 0.99) as a function of the MEKAL temperature and the column 
  density of the interstellar absorption. The bright phase  interval was selected individually 
  for each of the five orbits. Note the different scaling of the y axis compared to Fig. \ref{x_spec_fit}.}
\label{x_brightspec_fit}
\end{figure}

We also created a combined spectrum for the bright phase intervals
before the absorption dip occurs. This corresponds to the orbital
phase interval of $\sim$0.6--0.8.  The phase intervals were visually
selected for each individual orbit.  A fit with an absorbed MEKAL
model resulted in a plasma temperature of kT = 14(4) keV and a column
density of $N_{\rm{H}} = 2.1(7) \times 10^{20}$ cm$^{-2}$
($\chi^2_{\nu}$ = 1.01). The spectral fit together with the confidence
range is shown in Fig.~\ref{x_brightspec_fit}. Since at the given
phase interval the accretion region is directly exposed towards the
observer, a fit to this spectrum reveals the most likely plasma
temperature. The unabsorbed flux is $2.8 \times
10^{-12}$\,erg\,s$^{-1}$\,cm$^{-2}$ in the energy range
0.15--12.0\,keV, the bolometric flux $3.8 \times
10^{-12}$\,erg\,s$^{-1}$\,cm$^{-2}$ that results in an accretion
luminosity of L$_{accr} \ge 0.55 \times 10^{32} d^2_{\rm{350}}$\,erg\,s$^{-1}$,
where $d_{\rm{350}}$ is the distance in units of 350 pc.
The luminosity can be
compared to the expected accretion luminosity. According to the
standard theory, the mass transfer below the period gap is driven
solely by the angular momentum loss due to gravitational
radiation. The resulting mass transfer rate for the derived orbital
period would be $\dot{M} = 2.8 \times 10^{-11}$ M$_{\odot}$ yr$^{-1}$
\citep{1995cvs..book.....W} and lead to an accretion luminosity of
L$_{acc} = 1.6 \times 10^{32}$\,erg\,s$^{-1}$ (for the used masses of
white dwarf and secondary see Sect.~\ref{binsys}). This is in good
agreement with the actually measured luminosity.

Contrary to the general assumption of accretion in AM Her stars, which
emerged from the analysis of the large body of e.g.~ROSAT data, the
X-ray spectrum of this object shows no evidence for the presence of a
distinct soft component. In most AM Her stars, the soft component
originates from the heated atmosphere of the white dwarf around the
accretion spot. Thus, \xmmp joins the group of now 9 out of $\sim$80
polars without a soft component \citep{2007MNRAS.379.1209R}.  A likely
explanation for the non-detection of the soft component is its low
temperature which shifts the emission towards lower energies not
covered by the X-ray cameras of the XMM-Newton. To derive upper limits
for the soft component, we model this component with a black body
spectrum based on the constraints that the Rayleigh-Jeans tail of the
assumed black body must not exceed the observed spectral flux in the
ultraviolet, and that the contribution of the soft component to the
0.15--0.30\,keV band is less than 10\%.  With a 0.6 \msun white dwarf
and a distance of 900 pc, the fraction of the white dwarf surface that
is covered by the pole cap falls below $10^{-3}$ for a temperature
above 15 eV that is considered as the upper temperature limit.

This temperature is at the lower end of the otherwise observed
temperatures for the soft component, but is still too high to
contribute in the UV. To test the energy balance of the reprocessed
component we assumed a black body with a temperature of 10 eV,
providing 10\% of the flux in the 0.15--0.30\,keV band (see
Fig. \ref{sed}).  The resulting contribution in the UVW2 filter is
still less than one fourth, but the integrated flux exceeds the flux
of the bremsstrahlung component by a factor of hundred. Thus, the UV
and X-ray observations give no constraints for the maximum flux
contribution of the soft component that could lead to a conflict with
the standard model.  In summary, the missing soft component is not
really puzzling. If the accretion occurs over a large fraction of the
white dwarf surface, the temperature of the pole cap would be lower
than observed in polars with soft component. The lower temperature
simply shifts the observable soft component towards the EUV, out of
the XMM band pass.
\subsection{The X-ray eclipse emission}
\label{eclipseemission}
The EPIC light curve shows a residual X-ray flux of $F_{X} = 1.2
\times 10^{-14}$\,erg\,cm$^{-2}$\,s$^{-1}$ when the white dwarf and
the accretion region are eclipsed by the secondary.  The origin of
this emission is not obvious at first glance.  \xmmp\, is accompanied
by SDSS J131223.75+173701.2 at a distance of 4.4 arcsec.  The
spectroscopic observations show that this object is a quasar at
redshift $z = 2.43$.  The quasar is not resolved in our XMM EPIC data,
although quasars are known to be X-ray emitters. We tried two methods
to distinguish between the two sources to clarify if the residual
X-ray emission emerges from the AGN or from the corona of the
secondary star (if the secondary is X-ray active at all).

Our first attempt was to spatially resolve the polar and AGN by using
phase selected event lists for bright, faint and eclipse phase on
which the SAS source detection was run. All three source positions
found were the same within the error circles, which is not surprising
since the distance between both sources is at the limit of the spatial
resolution.

The next attempt was to use the properties of the spectral energy
distribution.  We computed the flux from the dozen photons during the
eclipse, taking the interstellar absorption from the spectral fit
above into account. The X-ray flux of $F_{\rm{X}} = 1.2 \times
10^{-14}$\,erg\,cm$^{-2}$\,s$^{-1}$ was then compared with the optical
flux. From the SDSS magnitudes of the AGN we computed a optical flux
of $F_{\rm{opt}} = 2.76 \times 10^{-14}$\,erg\,cm$^{-2}$\,s$^{-1}$
\citep{1990hsaa.book.....Z}.  The ratio $F_{\rm{X}}$/$F_{\rm{opt}}$ =
0.4 can be compared with the properties of other AGNs. For most AGNs
$F_{\rm{X}}$/$F_{\rm{opt}}$ is found to be between 1 and 10, but can
also be one order of magnitude less, while for stars this ratio is
mostly below 0.05 \citep{2000AN....321....1S}.  Ascribing the complete
flux to the secondary, the X-ray flux transforms into a luminosity of
$L_{\rm{X}} = (1.8 - 11.6) \times 10^{29}$\,erg\,s$^{-1}$. The assumed
spectral type and the distances (see Sect.\ref{diffrband}) lead to a
bolometric luminosity of $L_{\rm{bol}} = (3.18 - 3.43) \times
10^{30}$\,erg\,s$^{-1}$ \citep{1996ApJS..104..117L}, and thus
$L_{\rm{X}}$/$L_{\rm{bol}} \sim$ 0.3 for an M5 and $\sim$ 0.06 for an
M6. This is orders of magnitudes above the value which could be
expected for an active M dwarf \citep{2003A&A...397..147P}. As the
assumed distance increases $L_{\rm{X}}$ and thus
$L_{\rm{X}}$/$L_{\rm{bol}}$ also increases, making the ratio even
larger.

Therefore, we suggest to ascribe the residual X-ray emission during
the eclipse to the AGN. Since the contribution from the AGN is rather
small and the real flux contribution is uncertain, we neglect the
contribution by the AGN for any further analysis of the X-ray
data. Since we have more flux during the faint phase than during the
time of the eclipse, there must be an additional contribution at this
time. Either the self-eclipse is not complete or we see bremsstrahlung
reflected from the accretion stream. Due to the low number of photons
and the merged contribution from the quasar and the polar we could not
make any viable conclusion from the hardness ratio.
\subsection{The optical spectrum}

The optical spectrum of \xmmp clearly reveals its nature as a CV
through the presence of a blue continuum with strong, superposed
emission lines of H, He{\sc I} and He{\sc II}. In general, the optical
spectra of CVs are a mixture of contributions by the photosphere of
the white dwarf; the heated pole cap and the accretion spot on the
white dwarf; cyclotron radiation from the cooling plasma and
recombination radiation from the accretion stream; and the illuminated
secondary. Some attempts to disentagle those components can be made by
utilising the shape of the eclipse, extracting and comparing different
bands, and by subtracting bright- and faint-phase spectra with the aim
to single out one of those components.  Apart from the eclipse proper
the light curves in BVR bands are modulated by $\sim$ 50\% and show
one bright hump with a colour-dependent phase of maximum emission. The
shift between the B and R band humps could be caused by beaming
properties of the cyclotron radiation that is more beamed at higher
harmonics (shorter wavelengths). The difference between the average
spectra in phase intervals 0.5--0.8 (brightest phase interval) and
0.2--0.4 (faint phase), however, displays just a smoothly varying blue
continuum and does not show any feature which could be associated with
a cyclotron harmonic hump. The more likely explanation is to assume a
different approach to explain the existence of the humps. The centre
of the B band bright phase corresponds to the centre of the X-ray
bright phase at phase $\sim$ 0.9. Thus, the B band hump is likely to
originate from the heated pole cap, while the R band hump can be
explained by cyclotron origin (see Sect.~\ref{binsys}). This implies a
cyclotron component which contributes in the red and infrared,
indicative of a rather low field strength of $\sim$ 10--20\,MG.
Unfortunately, there was no indication of Zeeman split lines in the
spectra that could be used to determine the magnetic field strength
directly.

As a rough guess for the contribution of the cyclotron component, we
computed the flux difference between the faint phase and the bright
phase. This clearly is an upper limit for the cyclotron flux since it
contains an undetermined fraction of atmospheric emission from the
accretion-heated spot. Nevertheless, if we regard the contribution of
$1.1 \times 10^{-13}$\,erg\,cm$^{-2}$\,s$^{-1}$ during the bright
phase as of pure cyclotron origin and apply a factor of 2 for
correction of beaming properties and unobserved cyclotron features,
one gets $F_{\rm{cyc}}/F_{\rm{X}} \leq 0.1$ that demonstrates the
prevalence of plasma cooling via bremsstrahlung over cyclotron cooling
and also indicates low field strength \citep{2004ASPC..315..187B}.

\subsection{Overall flux distribution}
\begin{figure}
\includegraphics[width=6.5cm,angle=-90,clip=]{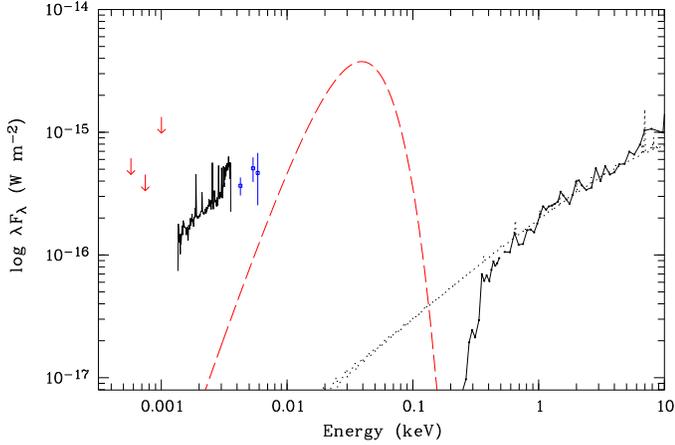}
\caption{Spectral energy distribution with all available data: 2MASS upper limits 
  (arrows on the left), CAHA mean spectrum (next to the arrows), XMM OM (next to the long dashed line) and XMM EPIC spectrum (on the right).
  Overplotted (long dash) is a 10 eV black body providing a negligible fraction of 10\% of 
  the observed soft X-ray flux (see Sect. 3.2 for a discussion of this component) and the
  MEKAL model spectrum (dotted line).}
\label{sed}
\end{figure}
The spectral energy distribution with all available data is shown in
Fig.~\ref{sed}.  \xmmp is not detected in 2MASS, thus an upper flux
limit could be derived for the JHK bands.  While the optical and the
X-ray spectra represent the mean over the orbital cycle, the UV data
points belong to specific orbital phases (see Table
\ref{2xmmom}). According to the spectral energy distribution, it is
clear that there is a strong flux contribution from the accretion
stream in the UV. The UVW1 filter covers just the faint phase, where
the accretion spot is self-eclipsed and does not contribute. Assuming
a low temperature white dwarf, the upper limit to the contribution of
the white dwarf is given by the optical flux, whereas assuming a high
temperature white dwarf, the upper limit is given by the flux in the
UVW2 band. A white dwarf model spectrum of T=20000 K could provide
$\sim$ 70 \% of the flux in the UVW1 band, and a T=11000 K white dwarf
$\sim$ 60 \%. The remaining flux has to be provided by the accretion
stream. To visualise the possible contribution of the soft component,
a black body spectrum of T=10 eV is also included in
Fig.~\ref{sed}. The black body provides 10\% of the X-ray flux in the
energy band 0.15--0.3 keV.
\subsection{Binary parameters and system geometry}
\label{binsys}
\begin{figure}
\centering
\resizebox{\hsize}{!}{
\includegraphics[bb=60 50 533 630, clip=]{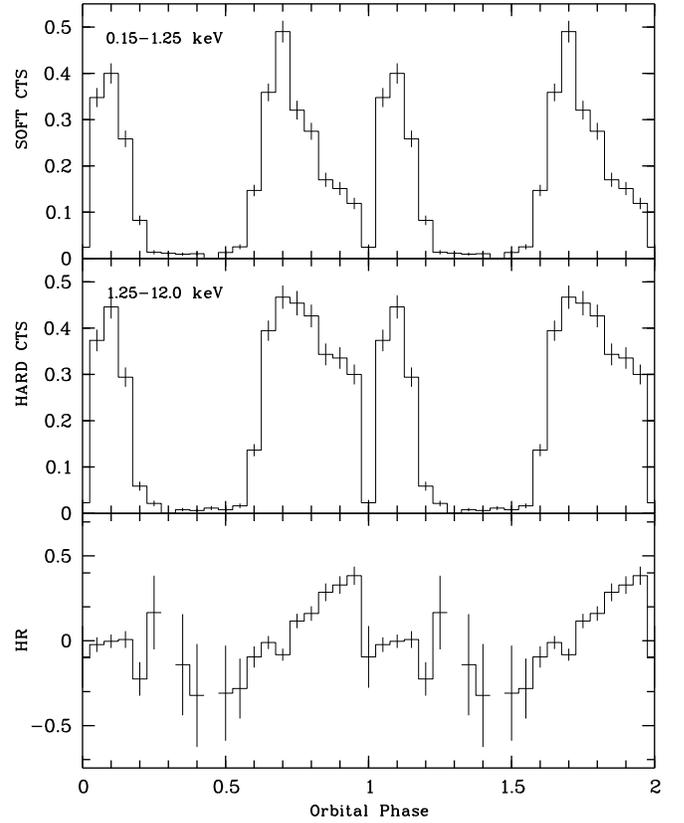}}
\caption{PN light curve for soft (0.15--1.25 keV) and hard (1.25--12.0 keV) energy ranges together with the resulting hardness ratio. Soft and hard band contain approximately an equal number of photons ($\sim$ 3100).}
\label{hr_curve}
\end{figure}
\begin{figure}
\includegraphics[width=6.5cm, bb = 42 54 568 792, angle = -90, clip=]{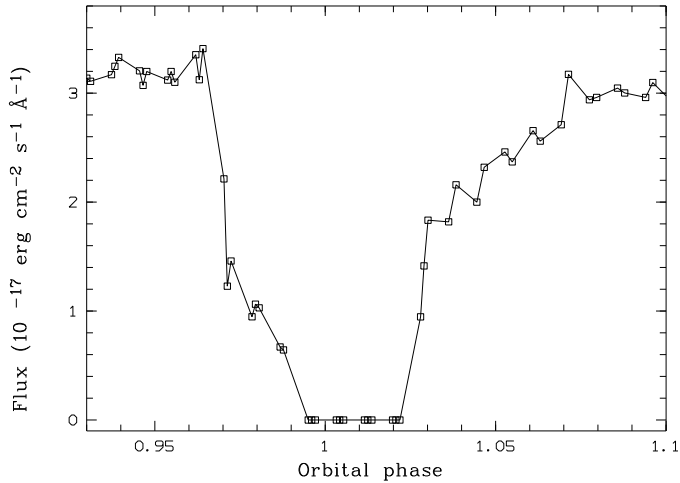}
\caption{Mean eclipse light curve from the optical R band observations on March~19 (errors
         are smaller than symbol size). Ingress and egress of accretion spot and accretion 
         stream are clearly identifiable. Non-detections are set to zero.}
\label{caha_stream_ecl}
\end{figure}

The optical mean eclipse light curve in Fig.~\ref{caha_stream_ecl}
indicates the ingress and egress of the white dwarf/accretion spot and
the stream. The width of the optical eclipse is 320$\pm$3 seconds
compared to the X-ray eclipse length of 329$\pm$3 seconds. The stream
ingress lasts until phase 0.995 and the stream egress until phase
$\sim$ 1.08. We used the steep decline during the eclipse ingress as
a consistency check for the binary distance as derived from the
spectral type of the secondary.  While the gradual decline is caused
by the accretion stream, the steep decline represents the eclipse of
the white dwarf and/or spot. Thus, the flux provided by the white
dwarf alone must not exceed the step height of $\sim 1.7 \times
10^{-17}$\,erg\,s$^{-1}$\,cm$^{-2}$\,\AA$^{-1}$.  With a mean white
dwarf mass of M$_1\sim$0.6 \msun\, and a mean temperature of 16000 K
\citep{1999PASP..111..532S}, the implied distance is d $\ge$ 550 pc,
which is consistent with the distances derived in
Sect.~\ref{diffrband}.

Assuming the secondary fills its Roche-lobe and by using the M-R
relation from \cite{2006MNRAS.373..484K} together with the period, we
get a secondary mass of M$_2\sim$ 0.09, which is only weakly dependent
on the mass ratio $q$. The mass of the secondary and the assumed white
dwarf mass of M$_1\sim$ 0.6 yields a mass ratio of $q\sim$ 0.15.  The
mass ratio, the orbital period, and the eclipse width determine the
orbital inlination. Since the white dwarf is not resolved in our data,
we used the mean spot eclipse width of 325 s to compute a $q-i$
relation that results at the inclination of $i\sim$ 83$^{\circ}$.
Since the real white dwarf mass remains unknown, the derived
inclination is just the approach of least prejudice, but hardly
constrains the system geometry.

The X-ray bright phase is centred at phase 0.9 that gives the azimuth
of approximately 36$^{\circ}$.  The length of the X-ray bright phase
is about 0.62 phase units locating the accretion spot on the
hemisphere most oriented towards earth. Since the length of bright
phase depends on the inclination $i$ and colatitude $\beta$ of the
accretion region, we get a constraint for the colatitude. Neglecting
any height of the accretion column, the colatitude is $\beta \sim$
20$^{\circ}$ for $i \sim$ 83$^{\circ}$, but correspondingly larger for
a possible vertical or horizontal extent of the emission region.

Both the optical and X-ray eclipses are preceded by a dip. This is
also found in other polars and is caused by absorption of X-rays when
the accretion stream passes the line of sight to the emitting
accretion region (\citealt{1987MNRAS.226..867W},
\citealt{2001A&A...375..419S}) as is required if $i > \beta$. In this
case, we expect to see a change in the hardness ratio since the
photoelectric absorption cross-section is greater at lower
energies. The hardness ratio together with the soft and hard X-ray
light curve can be seen in Fig.~\ref{hr_curve}. Interestingly, the
hardness ratio shows a monotonic increase from phase 0.75 until the
beginning of the eclipse.  If this is just due to the photoelectric
absorption, the corresponding accretion stream has to be extended over
a large range in azimuth and is more properly an accretion curtain
than a well-defined stream. This could also be an explanation for the
non-detection of the soft component of reprocessed origin.  An
extended accretion curtain leads to a bigger and thus lower
temperature accretion region on account of the lower specific
accretion rate.  We searched for evidence of an extended accretion arc
from the X-ray light curve at eclipse ingress and egress. Ingress and
egress are safely shorter than 5\,s and likely shorter than 3\,s but
the low number of counts does not allow to derive stringent
limits. For comparison, in the similar system HU Aqr, ROSAT-observed
eclipse egress lasts 1.3\,s \citep{2001A&A...375..419S}.

To test whether the dip is caused by absorption of cold material, we
tried to fit the X-ray spectrum of the phase intervals where the dip
occurs. The plasma temperature and normalisation were fixed to the
fitting values of the bright phase spectrum before the absorption dip
(see Sect.~\ref{xspectrum}). To ascertain whether the spectral change
between the bright phase and the dip can be explained with additional
absorption, the hydrogen column was left as a free parameter. 
With these constraints, a reasonable fit with a total N$_H =0.41(4) \times 10^{22}$cm$^{-2}$
and $\chi^2_{\nu}$ = 2.2 was obtained.
\begin{figure}
\includegraphics[width=6.0cm,bb=112 44 562 711,angle=-90,clip=]{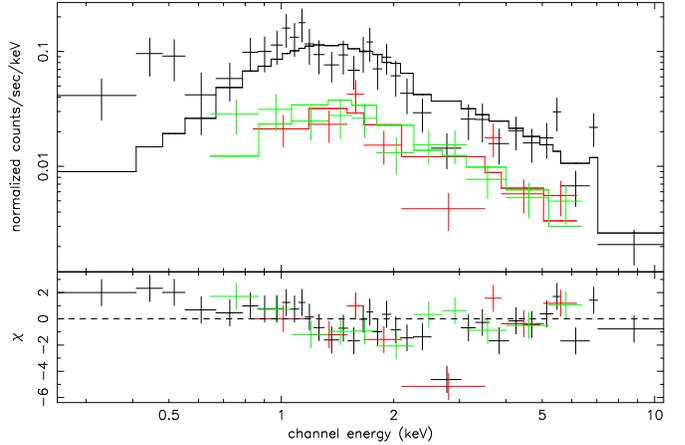}
\caption{Spectral fit of the combined PN and MOS spectra for the phase intervals of the 
        absorption dip (visually selected for each orbit) with an absorbed MEKAL model. Plasma
        temperature and normalisation were fixed to the fitting values of the bright phase spectrum 
        before the absorption dip.}
\label{dipfit}
\end{figure}

Assuming a high inclination as derived above, the centre of the stream
dip gives the azimuth of the threading region in the orbital plane.
The X-ray dip is centred at phase $\sim$ 0.9. This corresponds to an
azimuth of $\sim$ 54$^{\circ}$ of the threading region.  The optical
stream dip is centred at phase $\sim$ 0.85, somewhat before the X-ray
dip. The phase difference could either be due to a shifted spot
location or a higher accretion rate compared to the X-ray observation,
and thus an accretion stream coupling later in the orbital plane to
the magnetic field.

The centre of the R band hump occurs at phase $\sim$ 0.62, which is
0.3 phase units before the centre of X-ray bright phase. If we exclude
a radical change in accretion geometry between the X-ray observation
in 2004 and the optical observation in 2007, this can be explained by
cyclotron beaming, meaning a cyclotron origin of the R band hump.
With the spot location as derived above, the centre of the R band hump
corresponds to the orbital phase where the spot just appears from the
limb of the white dwarf. The optical hump starts at phase $\sim$
0.35, which is during the X-ray faint phase, and lasts until phase
$\sim$ 0.55. This suggests that the eclipse of the accretion column by
the white dwarf is complete only in X-rays and partial at optical
wavelength due to a high shock, so beamed cyclotron radiation is still
seen, while the X-ray emitting region is hidden behind the white
dwarf. The remaining puzzle is the missing second optical hump. With
this explanation a second hump would be expected, when the spot
disappears behind the limb of the white dwarf. The missing second hump
implies a rather large angle between magnetic field vector and surface
normal, which is hard to explain with a simple dipole geometry.

With the system parameters derived above, we modelled the extent of the
accretion curtain for the given system and binary parameters and were able to
reproduce the observed timings of the R band eclipse light curve. 
The long ingress phase of the curtain is due to the extended ballistic 
accretion stream which has its largest extent at an azimuth of $\sim$ 60\degr.  
\section{Summary}
We have presented the analysis of the newly discovered eclipsing
polar \xmmp.  The polar shows a hard X-ray spectrum with a plasma
temperature of $\sim$ 14 keV, but no sign of a soft component. The
soft component is probably shifted to the EUV due to an extended
accretion region and thus a cooler accretion region.  The system has
one active accretion spot on the hemisphere most oriented towards
earth with an azimuth of $\sim$ 36$^{\circ}$ and a low colatitude. The
orbital period, 92 min, could be determined with a fairly high
accuracy. The distance is likely to be $\ge$ 500 pc.

We could not directly determine the strength of the magnetic field, 
but the large extent of the ballistic stream and the properties of 
the cyclotron radiation suggest a rather low field strength of $B \leq 10$\,MG.
We suggest high-speed optical photometry observations with a large telescope to 
test our prediction of a large extended accretion 
arc and to determine the brightness and colour of the secondary in eclipse.
We further suggest IR spectroscopy and polarimetry to
directly determine the magnetic field strength. Finally, phase-resolved
UV photometry would help to determine the temperature of the white dwarf and the size 
of the accretion spot, allow to further disentangle radiation components, and 
settle the question of the undetected soft component.

\begin{acknowledgements}
JV and RSC are supported by the Deutsches Zentrum f\"ur Luft- und Raumfahrt
(DLR) GmbH under contract No. FKZ 50 OR 0404. KB acknowledges funding from
the European Commission under the Marie Curie Host Fellowship for Early Stage
Research Training SPARTAN, Contract No MEST-CT-2004-007512, University of
Leicester, UK. JO acknowledges the support of PPARC. We thank the German-Spanish
Astronomical Center at Calar Alto for allocating directors discretionary
observing time and the Calar Alto staff for competent execution of the service
mode observations.
\end{acknowledgements}

\bibliographystyle{aa}
\bibliography{references}
\end{document}